\documentclass[twocolumn,twoside,slac_two]{revtex4}
\usepackage{graphicx}
\usepackage{subfigure}
\usepackage{fancyhdr}
\usepackage{amsmath}
\usepackage{xspace}

\newcommand{\mett}{\mbox{${E\!\!\!\!/_T}$}}
\newcommand{\gt}{\mbox{$>$}}
\newcommand{\lt}{\mbox{$<$}}

\newcommand{\NONE}{\mbox{$\tilde{\chi}_1^0$}}

\newcommand{\HT}{H_{T}}

\newcommand{\etal}{{\em et al.}}

\newcommand{\DPHI}{\mbox{$\Delta\phi(\gamma_{1}, \gamma_{2})$}}
\def\Gravitino{\tilde{G}}
\def\bi{\begin{itemize}}
\def\ei{\end{itemize}}
\def\bc{\begin{center}}
\def\ec{\end{center}}
\def\and{\/\mbox{and}}

\newcommand{\grav}{\ensuremath{\tilde{G}}}
\newcommand{\gevc}{GeV/c$^2$}
\newcommand{\none}{\NONE}
\newcommand{\invfb}{$\rm fb^{-1}$} % Should be roman in PRL

\begin{document}

\title{Search for Supersymmetry Using Diphoton
Events in $p\bar{p}$ Collisions at $\sqrt{s}=1.96$~TeV}


\author{Eunsin Lee (for CDF Collaboration)}
\affiliation{Department of Physics, Texas A\&M University, College Station, TX
77843, USA}

\begin{abstract}
  We present the results of a search for supersymmetry with gauge-mediated
breaking
and $\NONE\rightarrow\gamma\Gravitino$ in
the $\gamma\gamma$+missing transverse energy final state. In 2.6$\pm$0.2~\invfb\
of $p{\bar p}$
collisions at $\sqrt{s}$$=$$1.96$~TeV recorded by the CDF~II detector we observe
no
candidate events, consistent with a standard model background expectation of
1.4$\pm$0.4
events. We set limits on the cross section at the 95\% C.L. and place the
world's best limit of 149~\gevc\ on the \none\ mass at
$\tau_{\tilde{\chi}_1^0}$$\ll$1~ns. We also exclude regions in the \none\
mass-lifetime plane for
$\tau_{\tilde{\chi}_1^0}$$\lesssim$$2$~ns.
\end{abstract}

\maketitle

\thispagestyle{fancy}


%%%%%%%%%%%%%%%%%%%%%%%%%%%%%%%%%%
\section{Introduction}\label{introtheory}
For theoretical reasons~\cite{gmsb}, and because of the
`$ee\gamma\gamma+$missing transverse energy ($\mett$)' candidate
event recorded
by the CDF detector in RUN I~\cite{run1evnt}, there is a compelling rationale
to search in high energy collisions for the production of heavy new particles
that decay producing the signature of $\gamma\gamma+\mett$.

An example of a theory that would produce such events is gauge
mediated supersymmetry breaking (GMSB)~\cite{gmsb} with
$\NONE\to\gamma\Gravitino$ where the $\NONE$ is the lightest neutralino and
the next-to-lightest supersymmetric particle
(NLSP) and the $\Gravitino$ is a gravitino which is the lightest supersymmetric
particle (LSP), giving rise to \mett\ by leaving the detector without
depositing any energy. The \grav\ also  provides a warm dark matter
candidate that is both consistent with inflation and astronomical
observations~\cite{astro}.

In these models, above the current limits from recent experiments~\cite{lep},
the \none\ is restricted to be well above 100~GeV and is favored to have a
lifetime on the order of a
nanosecond; the \grav\ is restricted to have a mass in the
range 0.5$<$${m}_{\grav}$$<$1.5~keV$/c^2$~\cite{cosmology}.
At the Tevatron sparticle production is predicted to be primarily into gaugino
pairs, and the \none\ mass ($m_{\tilde{\chi}_1^0}$) and lifetime
($\tau_{\tilde{\chi}_1^0}$) are the two
most important parameters in determining the final states and their
kinematics~\cite{gmsb}.
Depending on how many of the two $\NONE$'s
decay inside the detector, the event
has the signature $\gamma\gamma+\mett$, $\gamma+\mett$ or $\mett$ with one or
more additional high $E_{T}$ particles from the other gaugino pairs.
Different search strategies are
required for \none\ lifetimes above and below about a
nanosecond~\cite{prospects}.
Previous searches have been performed for low lifetime models in
$\gamma\gamma+\mett$~\cite{minsuk,d0search} and nanosecond lifetime models in
the delayed
$\gamma+jet+\mett$~\cite{delayedPRLD,lep} final
state.

In this analysis we focus on the $\gamma\gamma+\mett$ final state, as
recommended
in~\cite{prospects}, for low lifetime, high-mass models of the $\NONE$.
The new features of our analysis since the last $\gamma\gamma+\mett$ search
with 202 pb$^{-1}$ using the CDF detector~\cite{CDFII} are to use
the EMTiming system~\cite{nim} and a new {\it Met
Resolution Model}~\cite{ggXPRD}. We also use 13 times the data (2.6~fb$^{-1}$).
These additions significantly enhance our rejection
of backgrounds from instrumental and non-collision sources, which allows us
to considerably extend the sensitivity of the search for large \none\ masses
compared to previous Tevatron searches~\cite{d0search}. We also extend the
search by
considering \none\ lifetimes up to
2 ns which are favored for larger $m_{\tilde{\chi}_1^0}$.

Our analysis begins by defining a preselection sample by selecting events with
two isolated,
central \mbox{($|\eta|\lesssim 1.0$)} photons with $E_{T} > 13$~GeV.
All candidates are required to pass the standard CDF diphoton triggers, global
event selection, standard photon ID, and non-collision background rejection
requirements~\cite{minsuk,ggXPRD}.

The final signal region for this analysis is defined by the subsample of
preselection events that also pass a set of optimized final kinematic
requirements.
The methods for determining the background in the signal region are based on a
combination of data and Monte Carlo (MC) simulation and allow for a large
variety of potential final sets
of kinematic requirements.
We perform an {\it a priori}
analysis in the sense that we blind the signal region and select the
final event requirements based on the signal and background
expectations alone. We optimize our predicted sensitivity
using a simulation of our GMSB model. We then calculate, for each
GMSB parameter point, the lowest, expected 95\% C.L. cross section
limit in the no-signal scenario~\cite{limitcalc} as a function of the following
event variables:
MetSig, $\DPHI$, and $\HT$, each of which will be described in
Section~\ref{optimization}.

%%%%%%%%%%%%%%%%%%%%%%%%%%%%%%%%%%
\section{Data Selection}\label{dataset}
The analysis is based on 2.59$\pm$0.16 fb$^{-1}$ of
data delivered to the CDF detector in Run~II. The analysis selection begins with
events that pass the
CDF diphoton triggers which is effectively 100\% efficient for the final
diphoton selection requirements~\cite{ggXPRD}.
We require both highest-$E_{T}$ photons to be in the
fiducial part of the detector with
$|\eta|\leq 1.1$, pass the standard photon ID and isolation requirements and
have
$E_{T}^{\gamma}>$ 13 GeV. In addition to the standard photon ID requirements we
have
added additional requirements to suppress photomultiplier tube (PMT)
high-voltage
breakdowns (``spikes'')~\cite{delayedPRLD} and electron
rejection requirements~\cite{delayedPRLD} to
remove events where an electron fakes a prompt photon (Phoenix tracking
rejection). Each event is required to
have at least
one high quality vertex with $|z_{vx}|\leq$60 cm. The $E_{T}$ of all calorimeter
objects
(individual towers, photons, electrons, and jets) are calculated with respect to
the highest
$\sum P_{T}$ vertex.
However, an incorrect vertex can be selected when two or more collisions occur
in one beam-bunch
crossing, making it possible that the highest reconstructed $\sum P_{T}$
vertex does not produce the photons. If assigning the photons to a different
vertex lowers the $\mett$, we take that \mett\ and the photon $E_{T}$'s to be
from that vertex for all calculations (Vertex Re-assignment).

Additional standard selection requirements are placed to reduce
non-collision backgrounds, such as cosmic rays and beam-related (beam halo)
effects~\cite{delayedPRLD}.
We also apply $\mett$ quality
requirements (cleanup) to remove events if there is evidence that the
second photon ($\gamma_{2}$) or a jet is partially lost in a crack between detector
components~\cite{ggXPRD}.
Our pre-selection
sample consists of 38,053 events left after all the quality,
ID and cleanup requirements are applied~\cite{ggXPRD}.

%%%%%%%%%%%%%%%%%%%%%%%%%%%%%%%%%%
\section{Backgrounds}\label{background}
There are three major
sources of background for $\gamma\gamma+\mett$ events:
 QCD events with fake $\mett$, electroweak events with real
$\mett$, and non-collision events (PMT spikes, cosmic ray or beam-halo events
where
one or more of the photons and \mett\ are not related to the collision).

Standard Model QCD sources, $\gamma\gamma$,
$\gamma-jet\to\gamma\gamma_{fake}$, and $jet-jet\to\gamma_{fake}\gamma_{fake}$,
are the dominant producer of events in
the diphoton final state and a major background for $\gamma\gamma$ with fake
$\mett$.
These backgrounds
come in two different categories;
fake $\mett$ due to energy measurement fluctuations in the calorimeter and fake
$\mett$ due to pathologies such as
picking
the wrong vertex in events where the true collision did not create a vertex
or tri-photon events with a lost photon.

To estimate the background due to energy measurement fluctuations we use the
{\it Met Resolution Model}. The {\it Met Resolution Model} considers the
clustered and unclustered energy in
the event and calculates a probability,
$P({E\!\!\!\!/_{T}~\!\!\!\!^{fluct}}>\mett)$, for fluctuations in the energy
measurement to produce
${E\!\!\!\!/_{T}~\!\!\!\!^{fluct}}$ equivalent to or larger than the measured
$\mett$. This probability is then used to define MetSig~$=-$$\log _{10}
\left(P_{{E\!\!\!\!/_{T}~\!\!\!\!^{fluct}}>\mett}\right)$. Events with true and
fake $\mett$ of the same value should have, on average, different MetSig.
For each data event we throw 10 pseudo-experiments to generate
a $\mett$ and calculate its significance, according to the jets and underlying
event configuration. Then we count the number of events in the
pseudo-experiments that pass
our MetSig and other kinematic requirements. This number, divided by the number
of
pseudo-experiments, gives us the {\it Met Model} prediction for a sample.
The systematic uncertainty on the number of events above a MetSig cut
is evaluated by comparing the {\it Met Model}
predictions with the default set of model parameters to predictions obtained
with
the parameters deviated by $\pm\sigma$. The total uncertainty is estimated by
adding
the statistical uncertainty on the number of pseudo-experiments passing the cuts
and
these systematic uncertainties in quadrature.

A source of QCD background that is unaccounted for by the {\it Met Model} is
diphoton candidate events with event reconstruction pathologies such as a wrong
choice of
the primary interaction vertex or tri-photon events with a lost photon.
To obtain the prediction for the number of events with significant
reconstruction pathologies in the
QCD background at the same time, we model the kinematics and event
reconstruction  using a MC simulation of events with in the
detector using {\sc pythia}~\cite{pythia} and a {\sc geant}-based detector
simulation~\cite{geant}. We simulate a sample of SM
$\gamma\gamma$ events, with large statistics, and normalize
to the number of events in the presample to take into account jet backgrounds
which should have similar detector response.
Then we subtract off the expectations for energy measurement fluctuations in
the MC to avoid double counting. The remaining prediction is due to pathologies
alone.
The systematic uncertainties on this background prediction include the
uncertainty on the scale factor and the uncertainty due to MC-data differences
in the unclustered
energy parameterization and the jet energy scale.

Electroweak processes involving $W$'s and $Z$'s are the most common source of
real and significant $\mett$ in $p\bar{p}$ collisions. We estimate the
background rate from decays into both charged and neutral leptons using a
combination of data and MC methods.
There are four ways we can get a
$\gamma\gamma+\mett$ signature in electroweak events that decay into one or more
leptons: 1)~from $W\gamma\gamma$
and $Z\gamma\gamma$ events where both photons are real; 2)~from $W\gamma$ and
$Z\gamma$ events with a fake photon; 3)~from $W$ and
$Z$ events where both photon
candidates are fake photons; and 4) $t\bar{t}$ production and decay.
To estimate the contribution from the electroweak
backgrounds we use the Baur~\cite{baur} and {\sc pythia} MC's along with a
detector
simulation,
according to their production cross section and k-factors (the ratio of
the next-to-leading order (NLO) cross section to the leading order cross
section), but normalized to
data. 
To minimize the dependence of our predictions on potential ``MC-data''
differences, we normalize, using the rate of the number of $e\gamma$ events
observed
in the data that also pass
all signal kinematic cuts, to the number
of events observed in MC. This $e\gamma$ sample is derived from diphoton
trigger datasets and the events are required to pass the preselection
requirements where electrons are required to pass photon-like ID
requirements~\cite{ggXPRD}.
The uncertainty on the electroweak backgrounds are dominated by the
$e\gamma$ normalization factor uncertainty.
This includes data and MC statistical uncertainties
as well as differences in MC modeling.
The total uncertainties also include the MC statistical uncertainties
and uncertainties on the normalization factors added in quadrature.

Non-collision backgrounds coming from cosmic rays and
beam-related effects can produce  $\gamma\gamma$$+$$\mett$
candidates~\cite{delayedPRLD}.
These are estimated using the data.
Using the inclusive $\gamma\gamma$ sample selection requirements, but requiring
one of the photons to
have $t_{\gamma}$$>$$25$~ns we identify a cosmic-enhanced sample. Similarly, we
utilize a beam-related background enhanced sample. We
estimate the number of these events in the signal region using the ratio of
events outside the timing requirements to events inside the signal region
and the measured efficiencies of the non-collision rejection
requirements~\cite{ggXPRD}.
The uncertainties on both non-collision background estimates are dominated by
the statistical
uncertainty on the number of identified events. 

\begin{figure}[htbp]
  \centering
%   \subfigure[]{
%    \includegraphics[width=.48\linewidth]{../ggMet/cdfnote/pic/met_2.6fb_pub.eps}}
%   \subfigure[]{
 {   \includegraphics[width=1.\linewidth]{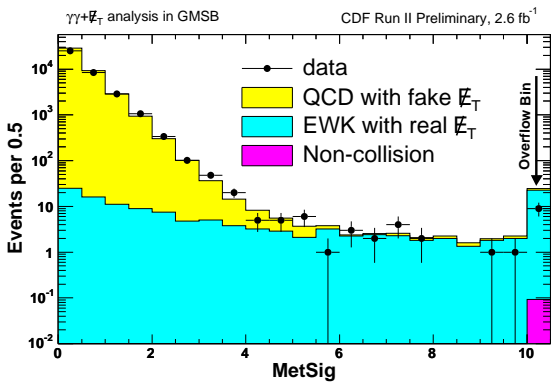}}
  \caption{The background predictions of MetSig for the
presample. The highest MetSig bin
includes all
overflow events.}
  \label{fig:bckdist}
\end{figure}

After estimating the MetSig distributions
for all the backgrounds, where the QCD is normalized to the data in the low
MetSig region where the EWK backgrounds are expected to be negligible, the
expected
MetSig distribution for the presample is shown in Figure~\ref{fig:bckdist}.
With these tools in hand
we are set to estimate the backgrounds for a large variety of kinematic
requirements and move to
an estimation of the acceptance for GMSB models in the signal region for use in
optimization.

\section{GMSB Signal Monte Carlo and Systematic Uncertainties}\label{acceptance}
To estimate the acceptance for GMSB we use the {\sc pythia} event
generator as well as a full detector simulation.
For the purpose of this analysis we consider a GMSB model with parameters fixed
on the
minimal-GMSB Snowmass slope constraint (SPS~8) that is commonly
used~\cite{minsuk,lep} and take the messenger mass
scale
  $M_{\mathrm{m}}$$=$$2$$\Lambda$, tan($\beta$)$=$15, $\mu$$>$$0$ and
  the number of messenger fields $N_{\mathrm{m}}$$=$1. The \grav\
  mass factor and the supersymmetry breaking scale $\Lambda$ are
  allowed to vary independently.
All SUSY production processes are simulated to maximize our sensitivity to the
model~\cite{simeon}.

Since we estimate the sensitivity of the search to be equal to the
expected 95\%~C.L. cross section limits with the no signal hypothesis, we need
the
uncertainties for the luminosity, background and acceptance. 
The systematic uncertainty on the luminosity
is taken to be 6\% with major contributions from the uncertainties on the CLC
acceptance from the precision of the detector simulation~\cite{geant} and the
event
generator~\cite{pythia}. 
The background uncertainty is
evaluated for every set of cuts in the optimization procedure.
The systematic uncertainty on the signal acceptance for an example GMSB point
of $m(\NONE)$~=~140~GeV and $\tau(\NONE)$~$\ll$~1~ns is estimated to be 6.9\%
with major contributions from diphoton ID and isolation efficiency (5.4\%) and
ISR/FSR (3.9\%).
The uncertainty on the NLO production cross section is dominated by the
uncertainty from parton distribution functions (7.6\%)
and the renormalization scale (2.6\%) for a total of 8.0\%.
All uncertainties are included in the final cross section limit calculation, and
we take the acceptance and production cross section uncertainties in quadrature
for a total uncertainty of 10.6\%.

\section{Optimization and Results}\label{optimization}
Now that the background is estimated and the signal acceptance is
available for a variety of selection requirements, an
optimization procedure can be readily employed to find the optimal
selection requirements before unblinding the signal region. We optimize for the
following kinematic requirements: MetSig, $\HT$, and $\DPHI$.

As described in earlier section, the MetSig cut gets rid of most of the
QCD
background with fake $\mett$.
The $\HT$ cut separates between the high $E_{T}$, light final state particles
produced by GMSB events via cascade decays and SM backgrounds, dominated by QCD
and electroweak backgrounds, which do not have lots of high $E_{T}$ objects.
The $\DPHI$ cut gets rid of events where
two photons are back to back since electroweak backgrounds with large $\HT$ are
typically a high $E_{T}$ photon
recoiling against
$W\to e\nu$, which means the gauge boson decay is highly boosted. Also the high
$E_{T}$
diphoton with large $\HT$ from QCD
background are mostly back-to-back with fake $\mett$ or wrong vertex.

\begin{figure*}[htbp]
  \centering
%  \subfigure[]{
%    \includegraphics[width=.46\linewidth]{../pic/sigmavsmetsig_2.6fb.eps}}
%  \subfigure[]{
%    \includegraphics[width=.50\linewidth]{../pic/n1_metsig_nodata_2.6fb_pub.eps}}
  \subfigure[]{
    \includegraphics[width=.46\linewidth]{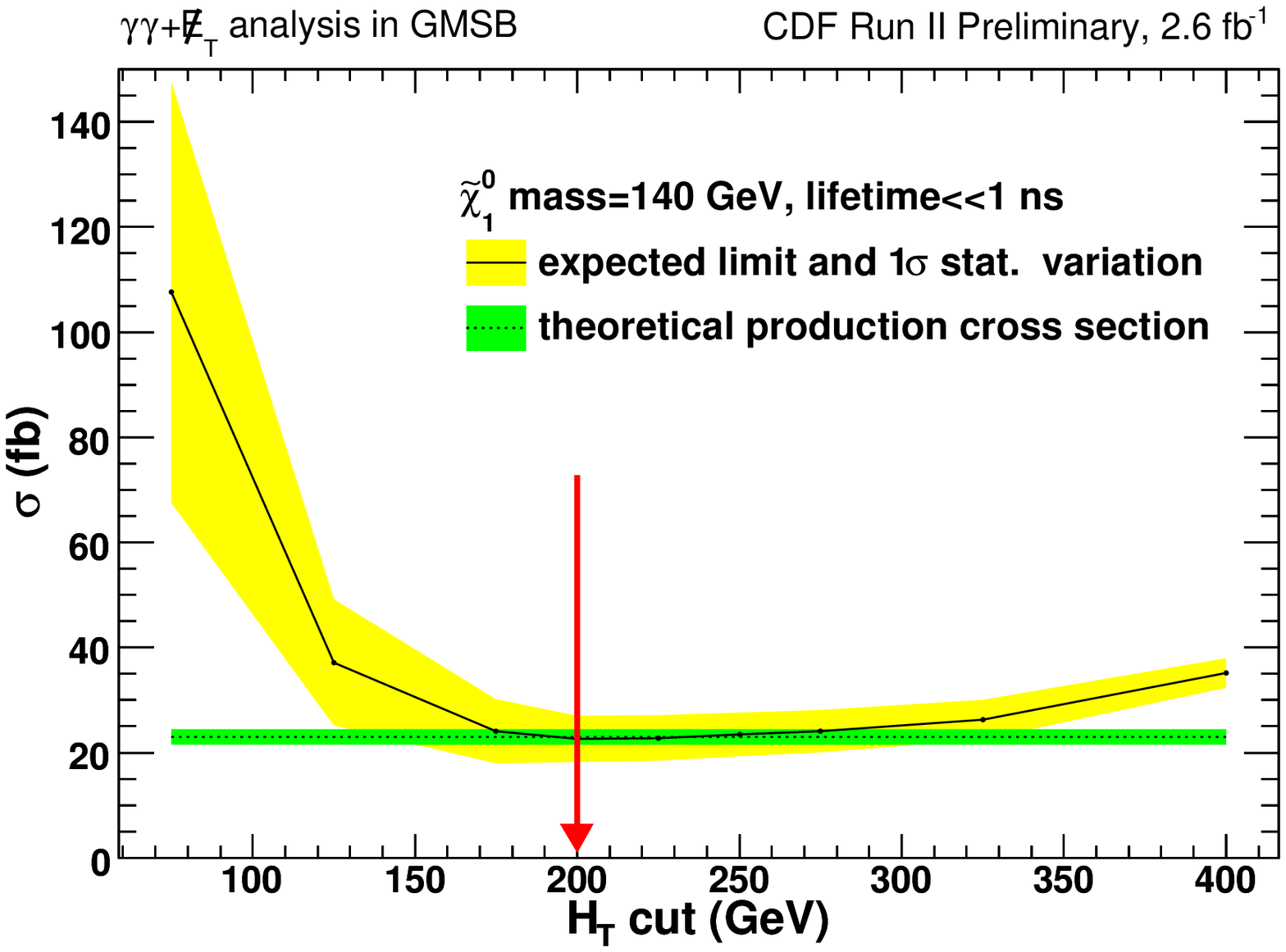}}
  \subfigure[]{
    \includegraphics[width=.50\linewidth]{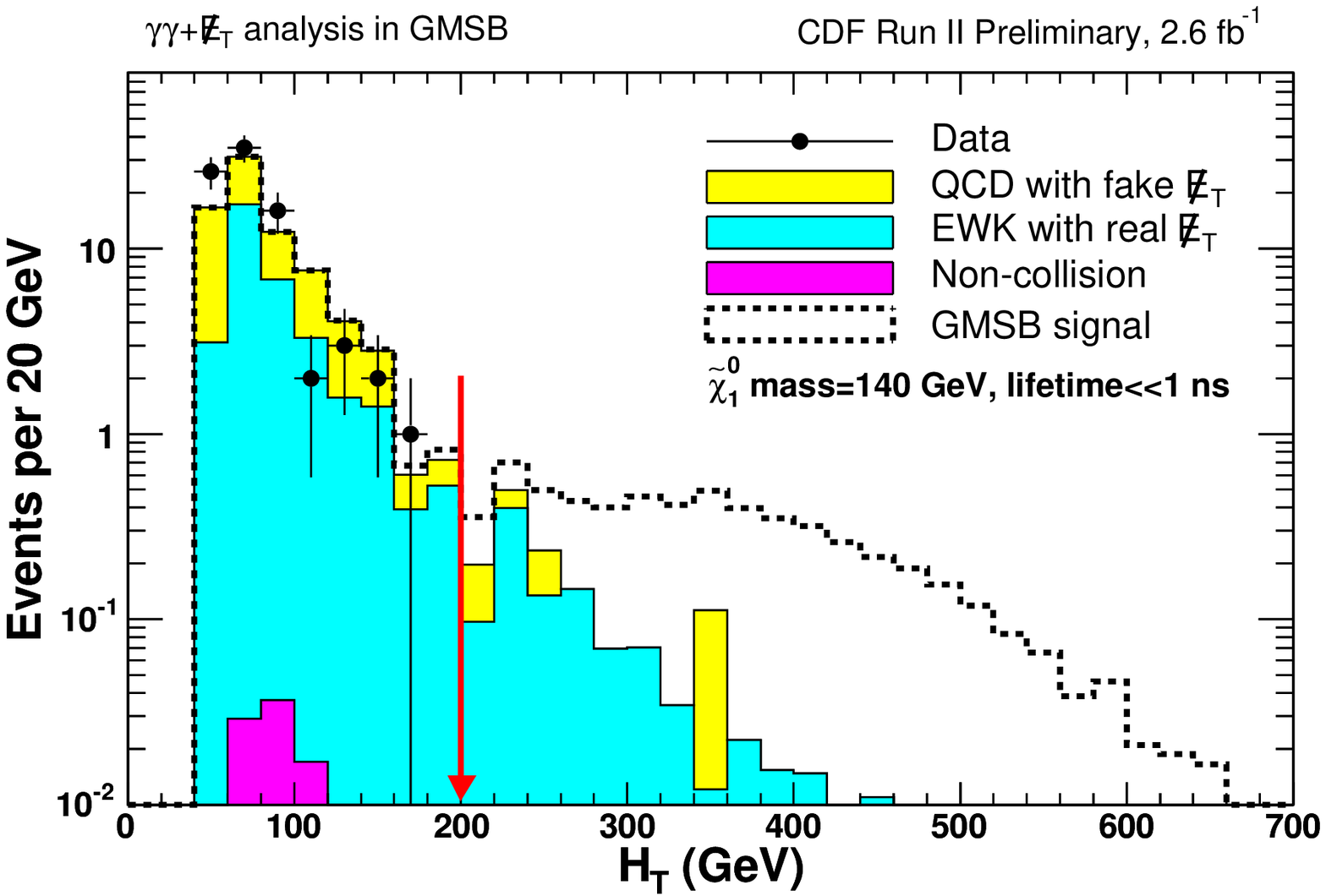}}
%  \subfigure[]{
%    \includegraphics[width=.48\linewidth]{../pic/sigmavsdphi_2.6fb.eps}}
%  \subfigure[]{
%\includegraphics[width=.50\linewidth]{../pic/n1_dphi_nodata_2.6fb_pub.eps}}
  \caption{The expected 95\% C.L. cross section limit as a function of
     the $\HT$~(a)
    requirement for a GMSB example point ($m(\NONE)=140$~GeV and
$\tau(\NONE)\ll1$~ns). All other cuts held at their optimized values.
    The optimal cut is where the expected cross
    section is minimized. Indicated in green is the 8.0\%
    uncertainty-band for the production cross section and in yellow is the RMS.
    The N-1 predicted kinematic distribution along with data after the optimized
requirements
 are shown in Figure (b). There is no evidence for new physics and the data
is well modeled by backgrounds alone.}
  \label{fig:xsection}
\end{figure*}

By estimating our sensitivity using the 95\%~C.L. expected cross section limits
on GMSB models in the no-signal assumption, we  find
the optimal set of cuts before unblinding the signal region.  We use the
standard CDF cross section limit calculator~\cite{junk} to
calculate the limits, taking into account the predicted number of
background events, the acceptance, the luminosity and their systematic
uncertainties.

For each GMSB point the minimum expected cross section limit defines our set of
optimal requirements for the mass and lifetime combination. The exclusion region
is
defined by the region where the production cross section is above the 95\% C.L.
cross section limit. The mass/lifetime limit is where the two cross.
Figure~\ref{fig:xsection}-(a) shows the expected cross section
limit
as a function of a kinematic selection requirement after keeping all other
requirements fixed at
the already optimized values, showing it is at the minimum for a mass-lifetime
combination of $m(\NONE)=140$~GeV and $\tau(\NONE)\ll1$~ns, which is near the
exclusion region limit.

We decided to use a single set of optimal requirements before we open the box
based on the observation that they will yield the largest expected
exclusion region. We chose: MetSig$\gt3$, $\HT\gt200$~GeV,
$\DPHI\lt\pi-0.35$~rad.
With these requirements we predict a total of 1.38$\pm$0.44 background events.
% with 0.77$\pm$0.30
%from electroweak
%sources with real $\mett$, 0.46$\pm$0.24 from QCD with fake $\mett$ and
%0.001$^{+0.008}_{-0.001}$ from non-collision.
The dominant electroweak
contributions are $Z\gamma\to\nu\nu\gamma$ and $Z\gamma\to\mu\mu\gamma$ which
produce a total of 0.26$\pm$0.08 and 0.19$\pm$0.10 events respectively.
The QCD background is dominated by
energy measurement fluctuations in the $\mett$, estimated using the {\it Met
Model}, to have a rate of 0.46$\pm$0.24 events.
The non-collision backgrounds are dominated by cosmic ray which have a rate of
0.001$^{+0.008}_{-0.001}$ events.

After all optimal cuts we open the box and observe no events, consistent with
the expectation of 1.2$\pm$0.4 events.
We show the kinematic distributions for the background and signal expectations
along with the data in Figure~\ref{fig:xsection}-(b). There is no distribution
that hints at an excess and the
data appears to be well modeled by the background prediction alone.

\begin{figure*}[htbp]
  \centering
  \subfigure[]{
    \includegraphics[width=.48\linewidth]{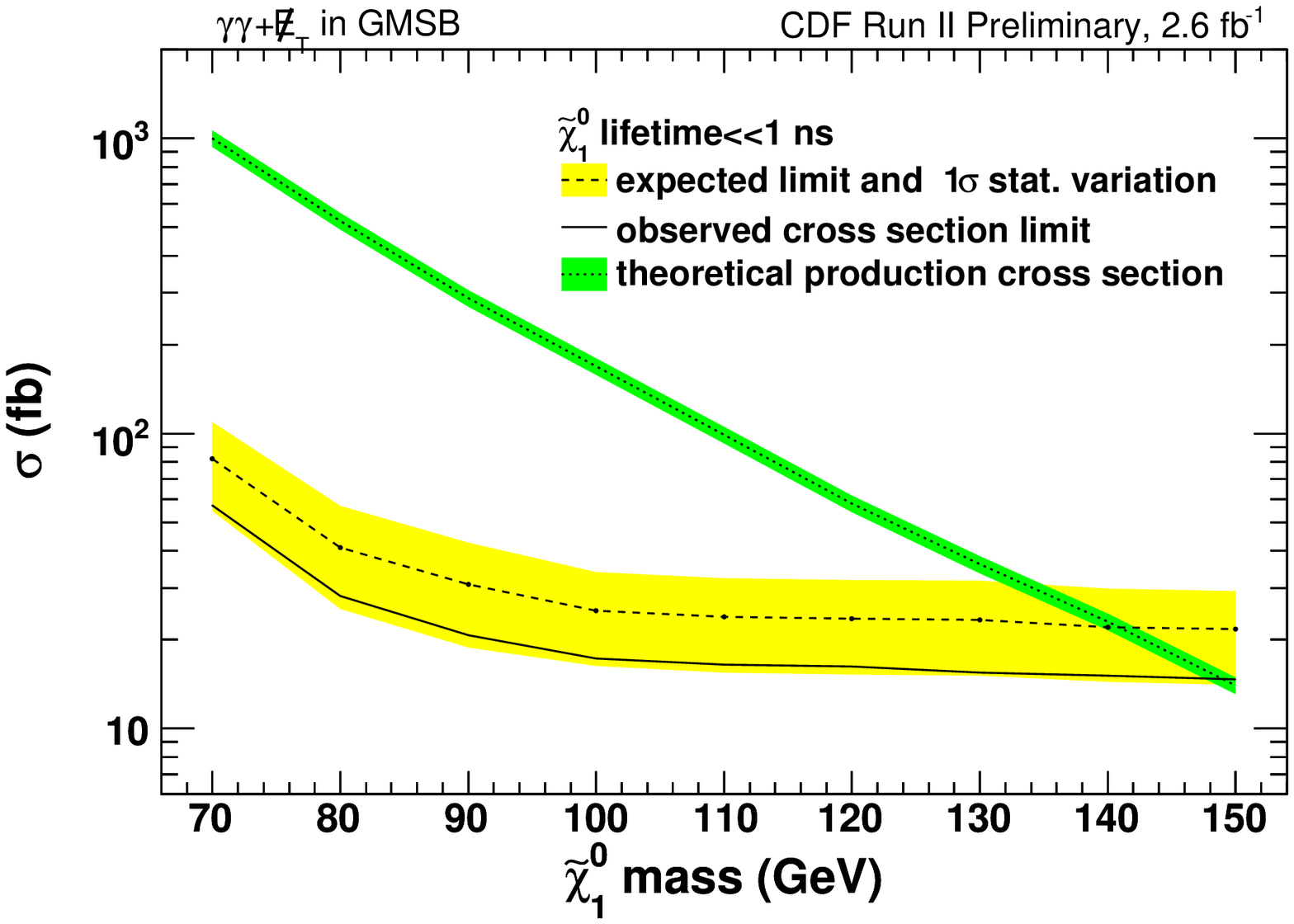}}
  \subfigure[]{
    \includegraphics[width=.48\linewidth]{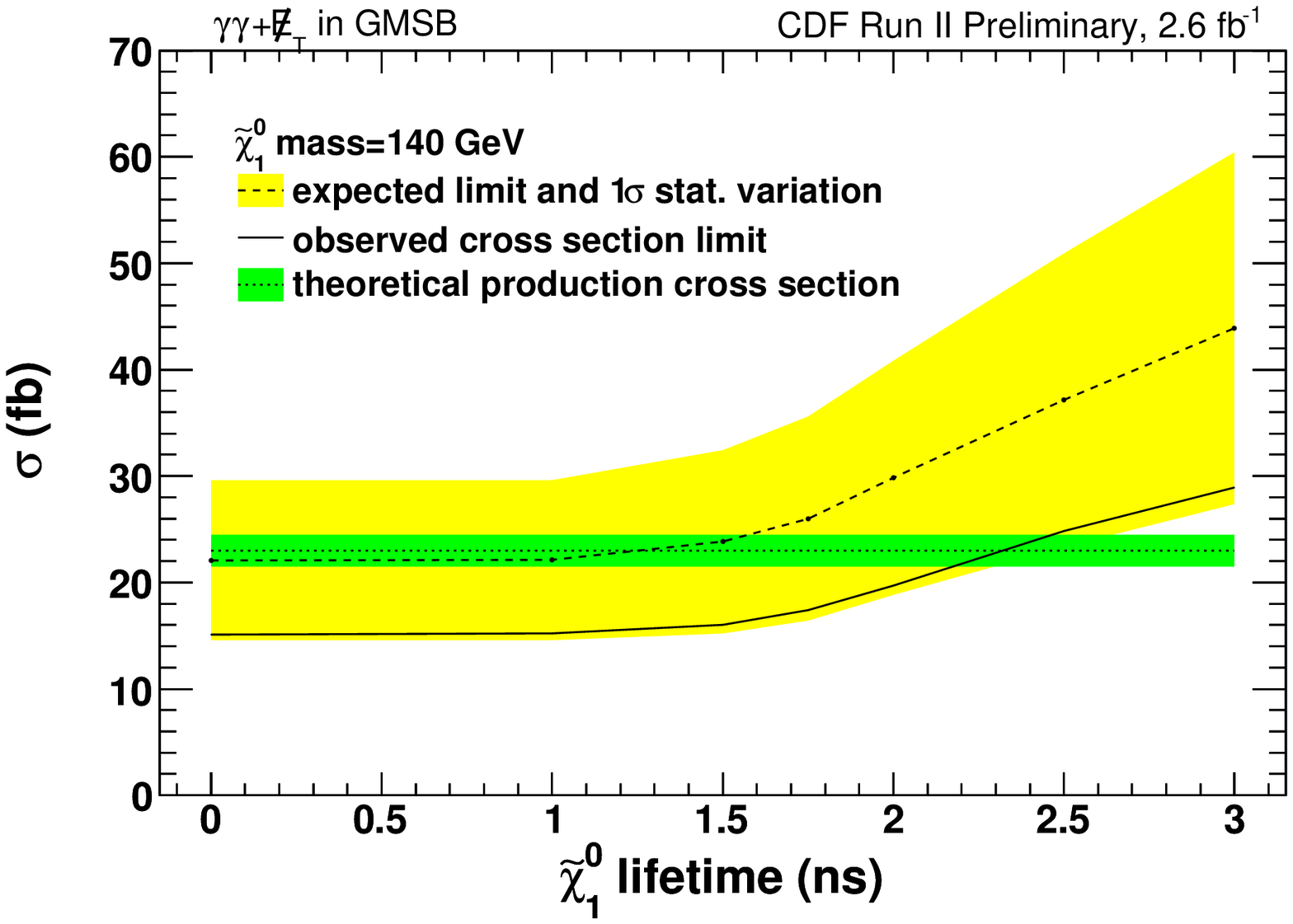}}
  \caption{The predicted and observed cross section limits as
    a function of the $\NONE$ mass at a lifetime much less than 1~ns (a) and as a
    function of the $\NONE$ lifetime at a mass of 140~GeV/$c^{2}$ (b). Indicated
    in green is the 8.0\% uncertainty-band for the production cross
    section, in yellow the RMS variation in the expected on the cross section
limit.}
  \label{fig:xsecmasslife}
\end{figure*}

\begin{figure*}[htbp]
  \centering
  \subfigure[]{
    \includegraphics[width=.51\linewidth]{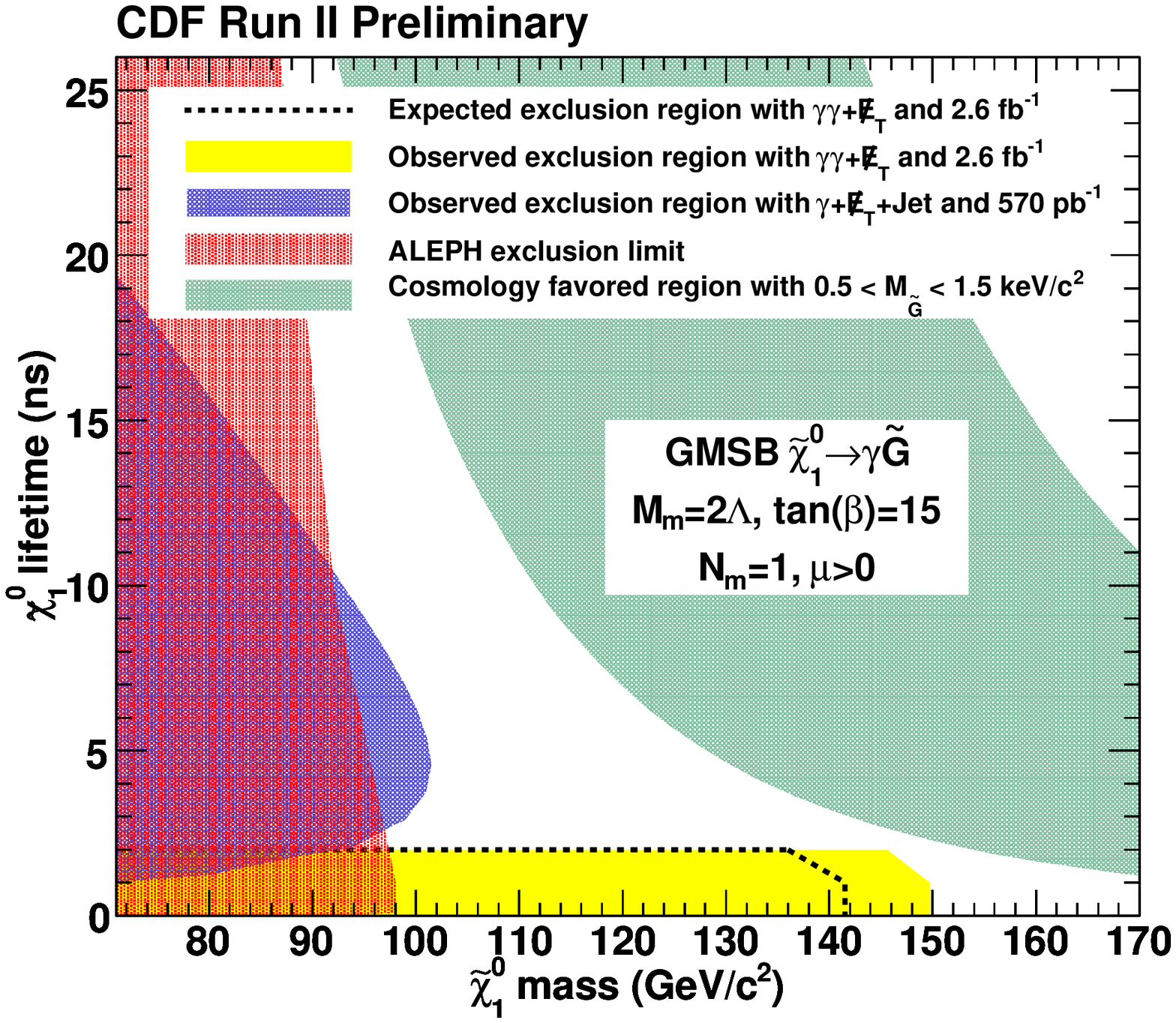}}
   \subfigure[]{
    \includegraphics[width=.45\linewidth]{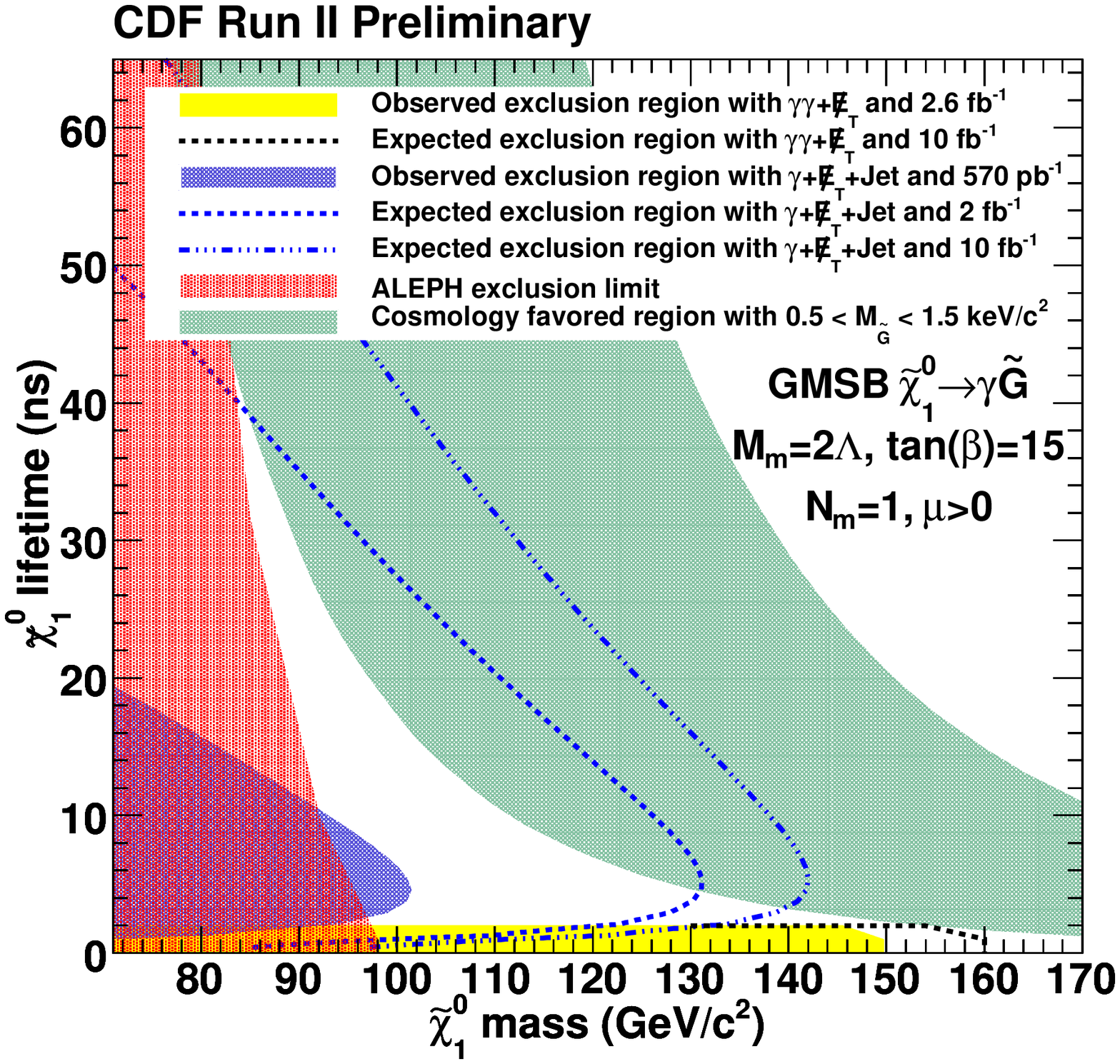}}
  \caption{The predicted and observed exclusion region along with the limit from
    ALEPH/LEP~\cite{lep} and the $\gamma+\mett+jet$ delayed photon
analysis~\cite{delayedPRLD}.
    We have a mass reach of 141~GeV/$c^{2}$
    (predicted) and 149~GeV/$c^{2}$ (observed) at the lifetime up to 1~ns. The
green shaded band shows the parameter space where $0.5<m_{\Gravitino}<1.5~{\rm
keV}/c^{2}$, favored in cosmologically consistent
models~\cite{cosmology} (a). The projected sensitivity to GMSB models with more
data.
The black dashed line shows the prediction of the exclusion region
limit after a scaling of the background prediction and the uncertainties for a
luminosity of 10~fb$^{-1}$. The blue dashed lines show the prediction of the
exclusion region limits from the delayed photon analysis for a luminosity of
2~fb$^{-1}$ and 10~fb$^{-1}$ respectively taken from Ref.~\cite{delayedPRLD}.
(b)}
  \label{fig:highlum}
\end{figure*}

We show the predicted and observed cross section
limits along with the NLO production cross section, which is calculated by
multiplying the {\sc pythia} LO cross section calculation by
k-factor~\cite{kfactors} in Figure~\ref{fig:xsecmasslife}.
Since the number of observed events is below expectations
the observed limits are slightly better than the expected limits. The $\NONE$
mass
reach, based on the predicted
(observed) number of events is 141~GeV/$c^{2}$ (149~GeV/$c^{2}$), at a lifetime
below 2~ns~\cite{prospects}.
We show the 95\% C.L. NLO exclusion region as a function of mass and lifetime of
$\NONE$
using the fixed choice of cuts from
the optimization for both for the predicted and observed number of
background events in Figure~\ref{fig:highlum}-(a).
These limits extend the reach beyond the CDF delayed photon
results~\cite{delayedPRLD} and well beyond those of D\O\
searches at $\tau_{\none}\ll0$~\cite{d0search} and the limit from
ALEPH/LEP~\cite{lep},
and are currently the world's best.

\section{Conclusions and Prospects for the future}\label{conclusion}

We have set limits on GMSB models using the $\gamma\gamma+\mett$ final state.
Candidate events were selected based on 13 times more data, the new \mett\
resolution model
technique, the EMTiming system and a full optimization procedure.
We found 0~events using 2.6~$fb^{-1}$ of data in run~II which is
consistent with the background estimate of 1.2$\pm$0.4 events from the Standard
Model expectations. We
showed exclusion regions and set limits on GMSB models with a $\NONE$
mass reach of 149~GeV/$c^{2}$ at a $\NONE$ lifetime much less than 1~ns. Our
results extend the world sensitivity to these models.

To investigate the prospects of a search at higher luminosity we calculate the
cross section limits assuming all backgrounds scale linearly with luminosity
while their uncertainty fractions remain constant. By the end of Run II, with an
integrated luminosity of 10~\invfb, we estimate a mass reach of $\simeq
160$~\gevc\ at a lifetime much less than 1~ns, as shown in Figure~\ref{fig:highlum}-(b).
For higher lifetimes (above $\sim$2~ns) the next generation delayed photon
analysis will extend the sensitivity taken from Ref.~\cite{delayedPRLD} and then
will combine these results for
completeness.

\begin{acknowledgments}
Eunsin Lee would like thank D.~Toback, R.~Culberton, A.~Pronko, M.~D'onofrio,
T.~Wright for their help on this analysis and the talk. 
\end{acknowledgments}

%%%%%%%%%%%%%%%%%%%%%%%%%%%%%%%%%%%%%%%%%%%%%%%%%%%%%%%%%%%%%%%%%%%%%%%%%%%
\end{document}